# Highly tunable valley polarization of potential-trapped moiré excitons in $WSe_2/WS_2$ heterojunctions

*Yueh-Chun Wu,[1,2] Matthew DeCapua[1,3] ZhongChen Xu,[4] Takashi Taniguchi,[5] Kenji Watanabe,[6] YouGuo Shi,[4] and Jun Yan[1,*]*

[1] Department of Physics, University of Massachusetts Amherst, Amherst, Massachusetts 01003, USA

[2] Materials Science and Technology Division, Oak Ridge National Laboratory, Oak Ridge, TN, 37831, USA

[3] Department of Materials Science and Engineering, Massachusetts Institute of Technology, Cambridge, Massachusetts 02139

[4] Beijing National Laboratory for Condensed Matter Physics and Institute of Physics, Chinese Academy of Sciences, Beijing 100190, China

[5] Research Center for Materials Nanoarchitectonics, National Institute for Materials Science, 1-1 Namiki, Tsukuba 305-0044, Japan

[6] Research Center for Electronic and Optical Materials, National Institute for Materials Science, 1-1 Namiki, Tsukuba 305-0044, Japan

[*]Corresponding Author: Jun Yan, yan@physics.umass.edu.

**ABSTRACT**

Moiré superlattices created by stacking atomic layers of transition metal dichalcogenide semiconductors have emerged as a class of fascinating artificial photonic and electronic materials. An appealing attribute of these structures is the inheritance of the valley degree of freedom from the constituent monolayers. Recent studies show evidence that the valley polarization of the moiré excitons is highly tunable. In heterojunctions of $WSe_2/WS_2$, marked improvement in valley polarization is observed by increasing optical excitation power, a behavior that is quite distinct from the monolayers, and lacks a clear understanding so far. In this work, we show that this highly tunable valley property arises from filling of the moiré superlattice, which provides an intriguing mechanism for engineering these quantum opto-valleytronic platforms. Our data further demonstrate that the long-range electron-hole exchange interaction, despite being significantly weakened in the junctions, is the dominant source of moiré exciton intervalley scattering at low population. Using magnetic field tuning, we quantitatively determine the exchange interaction strength to be 0.03 meV and 0.24 meV for 0° and 60° twisted samples respectively in our experiments, about one order of magnitude weaker than that in the monolayers.

In homo- and hetero-junctions of transition metal dichalcogenide (TMD) semiconductors twisted at angles near 0° or 60°, the interlayer coupling generates moiré patterns with a lattice constant much larger than that of the TMD monolayer, leading to a small moiré Brillouin zone allowing for efficient filling of quasiparticles in the moiré mini bands. This, together with the narrow moiré bandwidth compared to the Coulomb interaction strength in the system, has given rise to a plethora of correlated and topological phenomena[1–11]. TMD moiré superlattices also host rich and complex exciton physics [12–14]. In structures such as $WSe_2/WS_2$ and $WSe_2/MoSe_2$ heterojunctions, the type-II band alignment leads to optically active interlayer excitons where holes reside in $WSe_2$ and electrons reside in $WS_2$ or $MoSe_2$[15–18]. These excitons can be trapped in moiré cells where potential wells form from atomic registry dependent interactions between the TMD layers[19–28]. Each moiré cell can be viewed as a quantum dot hosting a quantum emitter[12], and single photon emission from moiré potential trapped individual interlayer exciton has been demonstrated experimentally[29,30]. Excitons in moiré cells also interact with each other. Recent studies have revealed biexciton quantum emitters in $WSe_2/MoSe_2$ and multi-exciton states in $WSe_2/WS_2$ heterobilayers, paving way for potential quantum information applications[21–23].

The valley degree of freedom is an important attribute in hexagonal TMD systems[31]. In TMD monolayers, the manipulation of the valley pseudospin played an important role in understanding the material system[32]. Stacked TMD heterojunctions inherit the valley quantum number from the monolayer (1L). The valley physics is, however, richer and more complicated due to involvement of different stackings, as well as formation of moiré superlattices and versatile atomic registries[26–28]. The interlayer exciton (IX) valley polarization (VP) can be highly sensitive to incident laser excitation power, a behavior not seen in 1L-TMDs, and which is so far not well-understood [11,23]. Several recent works have made use of the photoluminescence (PL) VP concept grafted from 1L-TMDs to explore interaction effects associated with correlated electronic phases at fractional fillings of the moiré minibands [10,11,33,34]. However, basic questions such as the atomic registry, spin and valley nature of the relevant moiré excitons, are controversial. Several contradictory assignments have been made for interlayer moiré exciton PL emission, especially for devices stacked near 60° [11,33,34].

In this Letter, we perform a comprehensive study on the spin and valley properties of neutral interlayer moiré excitons in $WSe_2/WS_2$ heterojunctions stacked at near 0° (R-type) and 60° (H-type). The detected optical emission exhibits robust photon helicity-valley index selection rules, similar to 1L-TMDs[31], allowing for definitive assignment of the spin and valley configuration, as well as the atomic registry for the moiré excitons. We confirm emission peaks associated with moiré cells occupied by one and two interlayer excitons, and they show distinct VP behavior in response to valley injection by circularly polarized optical excitation. In particular, the singly occupied moiré sites exhibit significantly improved exciton VP when the exciton generating rate reaches ~$10^{20}$ cm$^{-2}$ s$^{-1}$, initiating the population of the two-exciton state and facilitating the depletion of valley scattered single excitons. Intrinsic exchange interaction is found to provide the dominant intervalley scattering mechanism, like high-quality 1L-TMDs[35–39], despite its significantly weakened strength in moiré heterojunctions[14]. A marked improvement of moiré exciton valley polarization is observed upon application of a weak external magnetic field, effectively turning off this intervalley scattering channel. Our experiments provide a fairly direct measurement of the spin-triplet and spin-singlet moiré exciton exchange interaction energy $J_{ex}$, of ~ 0.03 and 0.24 meV for R- and H-type devices respectively, about an order of magnitude weaker than 1L-TMDs[36,40].

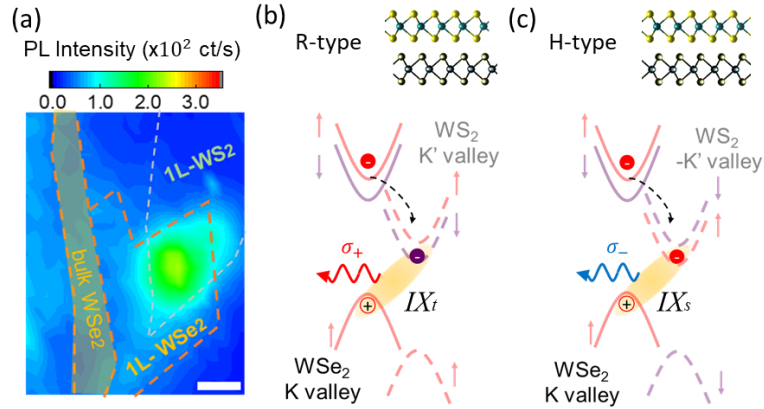

**FIG.1** (a) Interlayer exciton PL spatial mapping of a R-type WSe$_2$/WS$_2$ junction recorded with 532 nm, 0.2 mW excitation at 4 K. Scale bar: 2 μm. (b-c) Atomic stacking, band diagram, and ground state exciton optical emission in R-type (b) and H-type (c) WSe$_2$/WS$_2$ moiré junctions.

Our WSe$_2$/WS$_2$ heterojunction devices are fabricated using a dry transfer method and characterized by second harmonic generation[41] (see Supplemental Material[42]). Figure 1(a) shows typical photoluminescence (PL) mapping of an R-stacked sample. The interlayer exciton (IX) emission, due to radiative recombination of electrons in WS$_2$ and holes in WSe$_2$, appears in the junction area where monolayer PL is quenched (see Supplemental Material [42]). The nature of IX can be understood by considering the junction band configuration. In R-type (Fig.1(b)) and H-type (Fig.1(c)) devices, the K valley of WSe$_2$ is aligned with K' and -K' valleys of WS$_2$ respectively; except for a momentum displacement of about 0.5 nm$^{-1}$, due to the 4% lattice constant mismatch between WSe$_2$ and WS$_2$, corresponding to a moiré superlattice constant of ~8 nm. This leads to distinct spin configurations for ground state interlayer excitons in R- and H-junctions.

The exciton ground states, of spin-triplet ($IX_t$) in Fig.1(b) and spin-singlet ($IX_s$) in Fig.1(c), are responsible for IX PL emission in R- and H-type heterojunctions, respectively, according to our helicity resolved magneto-PL measurements in End Matter Fig.1. The $IX_t$ magnetic dipole moment is $4.5\mu_B$ (Zeeman splitting $\Delta E = E_K - E_{-K} = g\mu_B B$ with $g = -8.9$), of similar size to spin-triplet dark excitons in 1L-WSe$_2$ [43]. On the other hand, unlike the 0 out-of-plane angular momentum for 1L-WSe$_2$ dark excitons [43–45], $IX_t$ in the R-junction possess quantized angular momentum of $\hbar$ and $-\hbar$ in the two valleys, and emit σ+ and σ- circularly polarized PL from K and -K valley respectively. The brightening of $IX_t$ and its robust exciton valley - photon helicity connection is due in part to mirror symmetry breaking in the junction[28], and indicates that $IX_t$ inside the moiré cell is located in a region with a fairly uniform $C_3$ rotational symmetry, identified as a lattice-reconstruction expanded region with $R_h^X$ stacking (see Supplemental Material[42] for details). Interlayer exciton in H-junction has a magnetic dipole moment of $5.6\ \mu_B$ ($g = 11.2$, see End Matter Fig.1), consistent with the $IX_s$ assignment; the large $g$-factor arises from alignment of the K valley WSe$_2$ valence band with the -K' valley WS$_2$ conduction band, so that the plane-wave part of IX Bloch wavefunction contributes $\sim 4\mu_B$ to the magnetic dipole moment. $IX_s$ also has robust exciton valley index - photon helicity connection; interestingly, this IX carries an angular momentum of $-\hbar$ and emit photon with σ- polarization when its hole resides in the WSe$_2$ K valley, opposite to $IX_t$ in R-junction. From this flipped optical selection rule, we deduce that in H-moiré cell, the optical recombination occurs in regions with $H_h^h$ stacking symmetry (see Supplemental Material[42]). Note that prior studies have conflicting assignments on the location of moiré excitons inside the moiré cell, especially for the H-junction [11,33,34]. We comment that magneto-PL, such as those in our End Matter Fig.1, are more informative than zero field valley polarization measurements for making these assignments[16,33,34].

The robust exciton valley - photon helicity locking allows us to use the circular polarization of PL emission to assess the valley polarization of moiré excitons. Figure 2(a) shows helicity resolved PL of a R-stacked device excited by σ+ polarized, 730 nm wavelength light in resonance with the WSe$_2$ 1s exciton (see End Matter Fig.2 for H-stacked sample). The light was generated with a Ti:Sapphire tunable CW laser (M Squared SolsTiS), and the power incident on the sample was controlled using a combination of filters and polarizers. At a low incident power of 0.47 μW, the spectra are dominated by a single peak IX$_1$ slightly below 1.40eV that has a very small valley polarization. This mode is attributed to singly-occupied moiré-potential trapped *IX$_t$* illustrated in Fig.1(b). As the laser power increases, IX$_1$ valley polarization significantly improves, concomitant with the emergence of an additional mode IX$_2$ blueshifted from IX$_1$ by ~30 meV. This mode arises from a state with two excitons occupying one moiré cell (Fig.2d, lower).

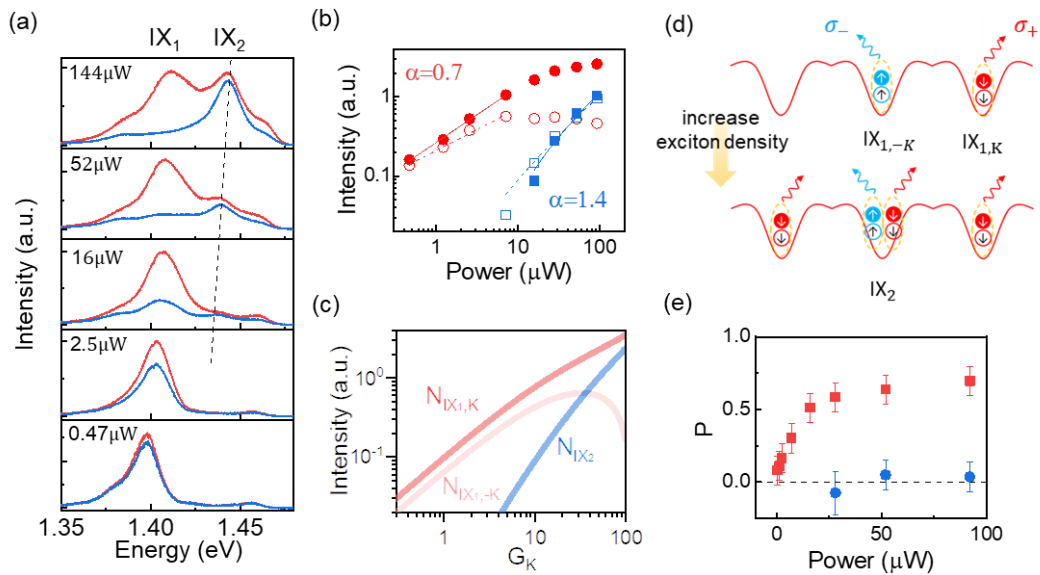

**FIG.2** (a) Helicity resolved (red: co-, blue: cross-), incident power dependent emission spectra of IX in a WSe$_2$/WS$_2$ moiré superlattice with R-stacking recorded with 730 nm excitation at 4 K. The two dominant peaks are assigned as moiré-potential trapped single-exciton (IX$_1$) and double-exciton (IX$_2$). (b) Log-log plot of IX$_1$ (red) and IX$_2$ (blue) intensity vs. incident power. Solid: co-circular polarization; open: cross-circular polarization. (c) Simulation of IX$_1$ and IX$_2$ population as a function of K valley IX$_1$ exciton generation rate G$_K$. (d) Illustration of IX$_1$ and IX$_2$ population as the incident power increases. (e) Emission polarization analysis for IX$_1$ (red) and IX$_2$ (blue).

The assignment of IX$_2$ to the two-exciton state is supported by its power dependence which scales roughly as the square that of IX$_1$; see Fig.2b, and agrees with other recent works[23]. The two-exciton state here is distinct from Coulomb-bound biexcitons free of potential confinement[46–48], where a net attractive binding is needed to stabilize the four-particle complex[49]. Here, optical emission from the moiré trapped two-exciton state has a higher energy than IX$_1$ due to on-site repulsion between the two excitons[23], reflecting the critical role of the trapping potential in the system[24]. The near 0 valley polarization of $IX_2$, together with the fact that it has almost the same Zeeman shift as $IX_1$ (End Matter Fig.1), indicates that it consists of a spin-one $IX_t$ in K valley, and a time-reversed spin-minus one $IX_t$ in -K valley located at the $R_h^X$ site of the moiré cell, as illustrated in Fig.2(d). Radiative recombination of an exciton in $IX_2$ has the same likelihood to originate from either valley, and the emission is thus unpolarized. We note that populating a moiré site with two excitons from two opposite valleys with opposite spins is energetically more favorable, due to the excessively large exchange interaction between two electrons or holes from the same valley with the same spin[23,40,50–53].

The $IX_1$ VP improvement with incident power is tightly linked to the spin anti-alignment of the two excitons in $IX_2$. The low VP of $IX_1$ at low power points to the presence of efficient valley depolarization processes when the moiré cells are populated with one-exciton states only. Meanwhile, the finite $IX_1$ VP at higher incident power suggests that circularly polarized σ+ optical excitation in resonance with the $WSe_2$ 1s exciton does generate valley-polarized excitons initially. To form a spin anti-aligned two-exciton state, an optically generated K valley IX needs to find a site that is already occupied with a -K valley IX. This process causes depletion of moiré sites occupied with a -K valley $IX_1$, and underlies the $IX_1$ VP improvement with increased optical excitation power.

To describe concretely the interplay between populations of $IX_1$ and $IX_2$, we developed a coupled rate equation model to understand the peculiar VP behavior of $IX_1$. Denote the generation rate of $IX_t$ excitons in K and -K valleys as $G_K$ and $G_{-K}$. At low incident powers without $IX_2$ formation, they are the sources for $IX_1$ population in the two valleys, $N_K$ and $N_{-K}$. At higher incident powers, the $IX_2$ population $N_{-KK}$ consists of two contributions: $\lambda G_K N_{-K}$ and $\lambda G_{-K} N_K$, where $\lambda$ describes the cross section for two-exciton state formation. The first term causes direct depletion of $N_{-K}$, and the second term decreases $G_{-K}$ that populates $N_{-K}$. Taking also into account finite exciton lifetime, the densities $N_K$, $N_{-K}$ and $N_{-KK}$ can be described by the following matrix formulation (see Supplementary Material[42] for details):

$$\begin{bmatrix} \gamma_K + \lambda G_{-K} & \lambda G_K & -\gamma_{-KK}/2 \\ \lambda G_{-K} & \gamma_{-K} + \lambda G_K & -\gamma_{-KK}/2 \\ \lambda G_{-K} & \lambda G_K & -\gamma_{-KK} \end{bmatrix} \begin{bmatrix} N_K \\ N_{-K} \\ N_{-KK} \end{bmatrix} = \begin{bmatrix} G_K \\ G_{-K} \\ 0 \end{bmatrix} \quad (1)$$

where $\gamma_K$, $\gamma_{-K}$ and $\gamma_{-KK}$ are the decay rates of $IX_1$ and $IX_2$. Note that annihilation of one exciton in $IX_2$ leaves behind an $IX_1$, forming an additional source term for $N_K$ and $N_{-K}$, which we have included in equation (1).

We solve for the behavior of our model quantitatively by using a normalized $\gamma_K = 1$, and $G_{-K} = 0.82 \times G_K$ that reflects a low-power VP of about 0.1 in Fig.2(a), where the excitons are well separated and IX$_2$ formation is negligible. Treating $G$ and $N$ as proxies for the incident power and emission intensity, we find that by setting $\lambda = 0.16$ and $\gamma_{-KK} = 12$ (IX$_2$ is reasonable to have a shorter lifetime than IX$_1$), our simulation in Fig.2(c) reproduces the behavior of data in Fig.2(b) fairly well.

It is of interest to link device physical quantities with the parameters used in the model. $\lambda$ is the cross section for a K valley IX to hit a moiré cell that contains a -K valley IX and form a two-exciton state. It should be of the size of a moiré cell, $10^{-13}$-$10^{-12}$ cm$^2$. Comparing data and model in Fig.2, $G_K = 10$ corresponds roughly to an incident power of 15 µW. Our incident photon is at WSe$_2$ 1s resonance where the absorption is roughly 15%,[54] from which we deduce at $G_K = 10$ the K valley exciton generating rate is ~$10^{20}$ cm$^{-2}$ s$^{-1}$. This then allows us to extract the exciton decay rate $\gamma_K = 10^7$-$10^8$ s$^{-1}$, in reasonable agreement with the measured interlayer exciton lifetime on the order of 100 ns.[55–57]

We now look into the microscopic processes of $G_K$ and $G_{-K}$ to assess what causes the significant IX valley depolarization at low power. In 1L-TMDs, there has been a fair amount of evidence indicating that exchange interaction plays a significant role in bright exciton valley depolarization[35–39], through the Maialle-Silva-Sham (MSS) mechanism[58]. For interlayer excitons, the exchange interaction is greatly weakened, especially for R stacked samples where the IX is a spin-triplet. It has been speculated that the MSS valley depolarization mechanism is strongly suppressed for the IX, as evidenced by their elongated valley lifetime[14]. Nonetheless, as we show below, the exchange interaction can still play a significant role in moiré IX valley depolarization.

We demonstrate this by analyzing the VP behavior in an external magnetic field. Figure 3(a) shows the magnetic field dependence of helicity resolved spectra for a R-stacked sample at a relatively low incident power of 2.7µW where two exciton formation is negligible. Interestingly, the IX PL emission becomes highly polarized upon application of a fairly weak magnetic field (about 120 mT) and switching $B$ direction does not switch the dominant circular polarization of the PL emission. This is at odds with VP improvement from majority population of the lower Zeeman mode[53]. Figure 3(b) plots the $B$ field dependence of $P = \frac{I_{\sigma+} - I_{\sigma-}}{I_{\sigma+} + I_{\sigma-}}$, from which we determine that the characteristic $B$ field (dashed red lines), that induces marked VP modulation, is $B_c^R \approx \pm(0.12 \pm 0.02)$ Tesla.

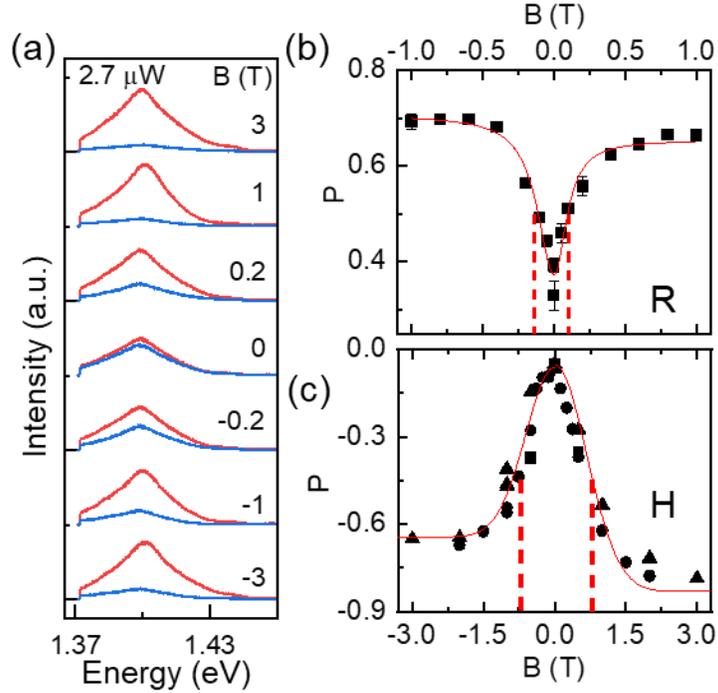

**FIG.3** (a) Helicity resolved magneto-PL spectra of an R-stacked sample recorded with 2.7 μW 730 nm excitation at 4 K. Red: co-circular; blue: cross-circular. (b) PL polarization vs. magnetic field for the R-staked sample. (c) Similar to (b) for a H-stacked sample (see spectra in End Matter Fig.2d). The different symbols correspond to measurements taken on different days with the same experimental parameters. Smooth curves are fits to $B$ dependence, and dashed lines indicate location of critical field according to the fits.

The significant VP enhancement in both positive and negative $B$ fields with the same sign indicates that the valley depolarization process is a resonant coupling effect that relies on the energy degeneracy between K and -K valleys. We attribute this coupling to the exciton exchange interactions, similar to observations for localized excitons in 1L-WSe$_2$[59]. The observed phenomena indicate that the IX exchange interaction strength is comparable to the exciton Zeeman shift at $B_c^R$, which is indeed much weaker than the exchange interaction of intralayer bright excitons.

Figure 4 illustrates the dominant IX formation processes. Our σ+ excitation in resonance with WSe$_2$ first creates K valley intralayer bright excitons. The spin up electron in WSe$_2$ subsequently tunnels to the WS$_2$ spin up conduction band. For H-type stacking, this is the only process that is needed to reach the IX energy ground state (Fig.4(b)) prior to radiative recombination. For R-type (Fig.4c), the electron experiences a subsequent spin flip (Fig.4(d)) to form the final $IX_t$ state.

Three types of exchange interactions (green dashes double arrows) are involved in Fig.4. $J_{ex1}$, the intralayer exchange, estimated to be of 0.5 - 2 meV [36–40], is too strong to account for the 0.12 T critical field observed in Fig.3(b). Distinguishing between $J_{ex2}$ and $J_{ex3}$ is more subtle since both are expected to be weak [27,28], although $J_{ex3}$ should be weaker since the exciton involved is a spin triplet. To our knowledge, there has not yet been a quantitative theoretical estimation or experimental determination of interlayer exciton exchange interactions.

Measuring similar $B$ dependence in H stacked samples provides useful insight. Figure 3(c) plots the $B$ dependence of $P$ for our H-type WSe$_2$-WS$_2$ junction (spectra in End Matter Fig.2(d)). In this case, we found that the Zeeman energy becomes comparable to exchange interaction at a field of $B_c^H \approx \pm(0.75\pm0.04)$ T. Since in H-type junction, formation of $IX_t$ is neither favored by energy relaxation nor by spin conservation (Fig.1c), we conclude that $J_{ex2}$ should be around 5.6 $\mu_B B_c^H \approx 0.24$ $meV$. This also allows us to deduce that in R-type junction, the dominant impact originates from $J_{ex3} \approx 4.5\ \mu_B B_c^R \approx 0.03$ $meV$. Our results thus point to eight times difference between $J_{ex2}$ and $J_{ex3}$. This is consistent with the fact that for spin-triplets, the valley depolarization process involves an additional step participated by spin-orbit interaction.

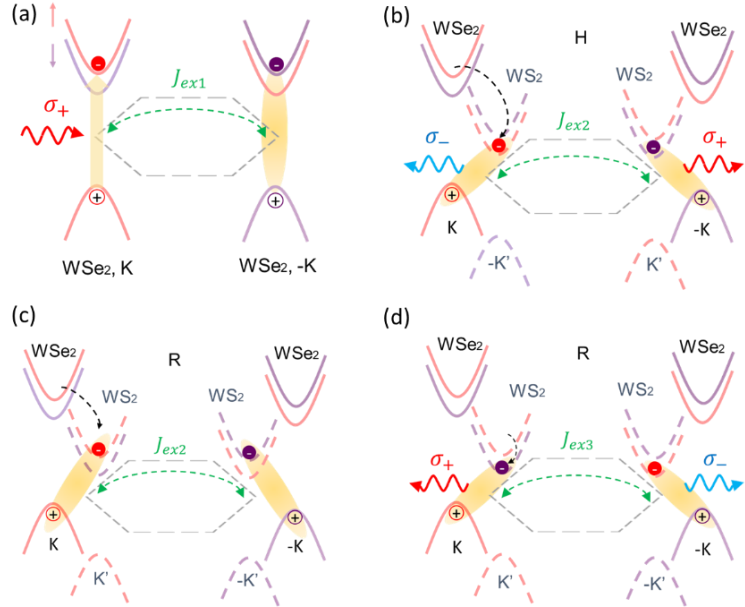

**FIG.4** Dominant exciton relaxation pathways and relevant exchange interactions. (a) Optical generation of WSe$_2$ K valley intralayer excitons. (b) Electron interlayer tunning to form $IX_s$ in H stacked device. (c) Similar electron tunning process in R-type sample. (d) Electron spin flip in R sample to form the $IX_t$ ground state.

We remark that the IX exchange interaction provides the dominant valley depolarization mechanism in the junction, as once it is turned off, VP reaches high values of ~70%. The residual depolarization could come from $J_{ex1}$. The high VP value indicates that the optically excited electrons stay in the WSe$_2$ layer only for an extremely short time[36], in agreement with the observed ultrafast interlayer charge transfer [60–62].

The insight we gain here is that the weakness of exchange interaction in TMD heterojunctions does not necessarily imply suppressed MSS-induced valley depolarization[58]. This is because exciton lifetime $\tau$ in the heterojunctions is much longer than that in monolayer[55–57], and $J_{ex}$ can act for a much longer time. In fact, since $\tau$ is inversely proportional to oscillator strength while $J_{ex}$ is proportional to oscillator strength, the Larmor precession, determined by $J_{ex}\tau$, is as effective for IX in the heterojunction as it is for 1s exciton in 1L-WSe$_2$. This is in contrast to 2s excitons in 1L-WSe$_2$, where the weak exchange interaction does lead to suppression of MSS and enhancement of 2s exciton valley polarization[36], as the 2s lifetime is shorter than 1s.

## ACKNOWLEDGEMENT

This work is supported by the National Science Foundation (DMR-2004474). K.W. and T.T. acknowledge support from the JSPS KAKENHI (Grant Numbers 20H00354 and 23H02052) and World Premier International Research Center Initiative (WPI), MEXT, Japan. Z.C. X and Y.G. S acknowledge support from the National Natural Science Foundation of China (Grants No. U22A6005) and the Synergetic Extreme Condition User Facility (SECUF). J.Y. is grateful for the hospitality of Westlake University where part of this work is done.

**End Matter:**

The IX optical emission in rhombohedral- (R-) and hexagonal- (H-) stacked samples are attributable to radiative recombination of spin-triplet excitons in Fig.1(b) located at $R_h^X$ and spin-singlet excitons in Fig.1(c) at $H_h^h$ inside a moiré cell, respectively (terminology following Ref.[28]). We made this assignment according to the helicity resolved magneto-PL data shown in End Matter Fig.1, where we compare IX emission with 1L-TMD PL.

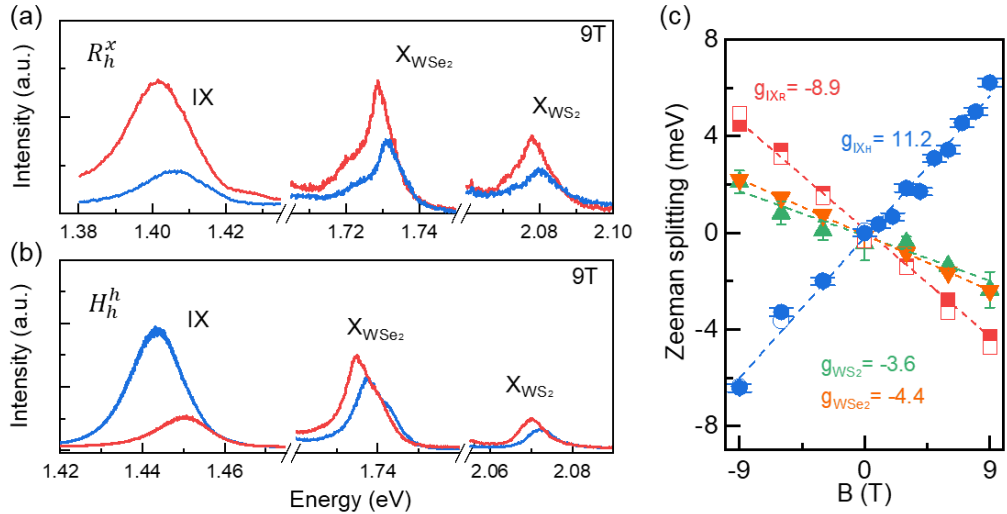

**End Matter Fig.1**: (a) Low-temperature (6K) magneto-PL of an R-junction. Red: σ+; blue: σ-. (b) Low-temperature (4K) magneto-PL of an H junction. The intralayer excitons were measured in the junction region for (a) over long integration, and measured in monolayer regions for (b). (c) Zeeman splitting of the exciton modes. Open symbols are extracted from IX$_2$.

In the presence of an external magnetic field, the Zeeman-split IX doublet in the two valleys radiate into σ+ and σ- emission channels respectively, reflecting quantized angular momentum of $\pm\hbar$. Note that while the lower Zeeman IX mode in R-junction has angular momentum $\hbar$ similar to the 1L-WSe$_2$ and 1L-WS$_2$, in H-junction it has $-\hbar$ angular momentum instead. This switching of angular momentum occurs when the K valley electron in WSe$_2$ tunnels into the -K' valley of WS$_2$ (Fig.4(b)), conserving spin but flipping the orbital angular momentum.

The Zeeman splitting between the modes in the two valleys, defined as $\Delta E = E_K - E_{-K} = g\mu_B B$, is plotted in panel (c) (the magnetic dipole moment for the valley excitons is $\pm\frac{g}{2}\mu_B$). The g values are -4.4, -3.6 for intralayer excitons, consistent with previous reports[63,64].

The g-factor of interlayer exciton in R-junction is -8.9, similar to spin-triplet dark excitons in WSe$_2$[43]. Taking into account the spin, atomic orbital, and valley (i.e. Bloch wavepacket) contributions to the magnetic dipole moment, we attribute the R-junction IX emission to spin-triplet excitons as illustrated in Fig.1(b), which has a lower energy than the spin-singlet, similar to monolayer WSe$_2$ and WS$_2$. The real space location of the IX in the moiré cell is determined from its spin-triplet nature and its measured angular momentum discussed above. According to Table S1 in Supplemental Material, the R-junction spin-triplet interlayer exciton is located at the $R_h^X$ area inside the moiré cell.

Similarly, the g-factor of 11.2 for H-junction PL is consistent with the magnetic dipole moment of the spin-singlet $IX_s$ in Fig.1c whose hole is in WSe$_2$ K valley and electron is in -K' valley of WS$_2$. With an angular momentum of $-\hbar$, we can look up Table S1 in Supplemental Material, and determine that the real space location of the H-junction $IX_s$ is $H_h^h$. Note that several prior works [11,33,34] used zero magnetic field valley polarization to deduce the location of the excitons inside the moiré cell. This is less reliable due to difficulties in distinguishing between intervalley scattering and atomic registry dependent optical selections rule, leading to conflicting assignments.

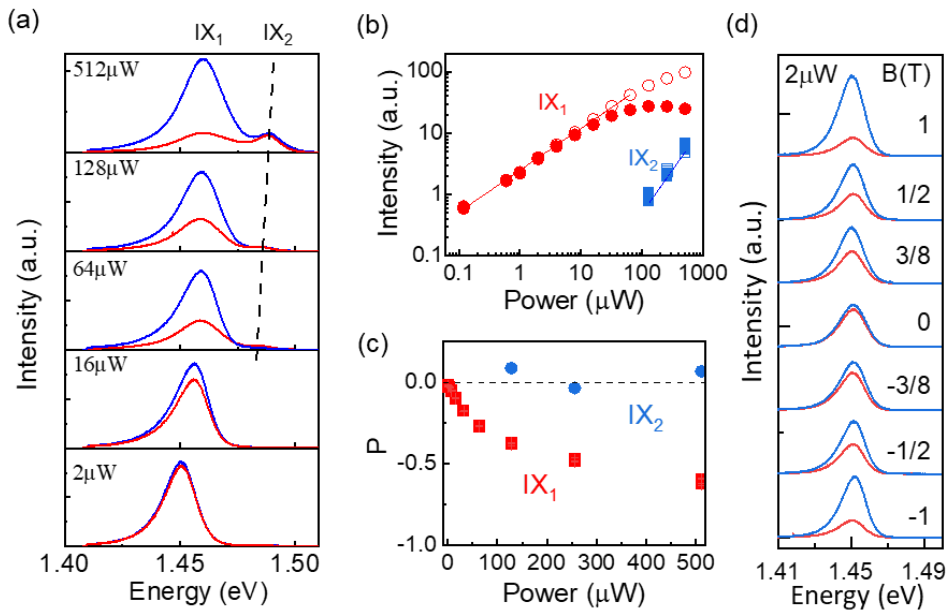

**End Matter Fig.2**: Data from H-junction, recorded with 730 nm excitation at 4 K. (a) Power dependence of helicity resolved PL emission. Blue: σ+σ-, Red: σ+σ+. (b) PL intensity of IX$_1$ (red circles) and IX$_2$ (blue squares). Open symbols: σ+σ-, closed symbols: σ+σ+. (c) Polarization $P = \frac{I_{\sigma+} - I_{\sigma-}}{I_{\sigma+} + I_{\sigma-}}$ of the PL emission. (d) Magnetic field dependence of the PL emission.

End Matter Fig.2 plots excitation power and magnetic field dependence of IX spectra on our H-junction sample, complementing the results of Fig.2 and Fig.3 in the main text for the R-type junction. In the power dependent spectra in panel (a), the VP of IX$_1$ increases rapidly with increasing laser power, while IX$_2$ has almost 0 VP throughout. Panels (b) and (c) plot the incident power dependent emission intensity and polarization, qualitatively similar to R-junction in Fig.2. The magnetic field dependence shows enhanced negative polarization for both positive and negative fields, indicating suppression of the exchange-interaction-induced valley depolarization channel by lifting the IX valley degeneracy, similar to Fig.3(a).